\documentclass[aps,twocolumn,showpacs,preprintnumbers,amsmath,amssymb,superscriptaddress,floatfix,nofootinbib,dvipdfmx]{revtex4}

\usepackage{graphicx}
\usepackage{epsfig}
\usepackage{epstopdf}
\usepackage{hyperref}
\usepackage{amsmath}
\usepackage{amsfonts}
\usepackage{amssymb}

\begin{document}

\title{\boldmath Semileptonic $\Lambda_c$ decay to $\nu l^+$ and $\Lambda(1405)$}
\date{\today}

\author{N. Ikeno} \email{ikeno@rs.tottori-u.ac.jp}
\affiliation{Department of Regional Environment, Tottori University,
  Tottori 680-8551, Japan}

\author{E.~Oset} \email{oset@ific.uv.es}
\affiliation{Departamento de
F\'{\i}sica Te\'orica and IFIC, Centro Mixto Universidad de
Valencia-CSIC Institutos de Investigaci\'on de Paterna, Aptdo.
22085, 46071 Valencia, Spain}

\begin{abstract}
 We study the semileptonic decay of $\Lambda_c$ to $\nu l^+$ and $\Lambda(1405)$, where the 
 $\Lambda(1405)$ is seen in the invariant mass distribution of $\pi \Sigma$. We perform the hadronization of the quarks produced in the reaction in order to have a meson baryon pair in the final state and then let these hadron pairs undergo final state interaction from where the 
$\Lambda(1405)$ is dynamically generated. The reaction is particularly suited to study this resonance because we show that it filters I=0. It is also free of tree level $\pi \Sigma$ production, which leads to a clean signal of the resonance with no background. This same feature has as a consequence that one populates the state of the  $\Lambda(1405)$ with higher mass around 1420 MeV, predicted by the chiral unitary approach. We make absolute predictions for the invariant mass distributions and find them within measurable range in present facilities. The implementation of this reaction would allow us to gain insight into the existence of the predicted two  $\Lambda(1405)$ states and their nature as  molecular states. 

\end{abstract}

\maketitle

\section{Introduction}

Semileptonic decays of the $\Lambda_c$ account for a fair fraction of
the total width. The branching ratios for decay into $\nu_e e^+ \Lambda$
and $\nu_{\mu} \mu^+ \Lambda$ are both about 2\%
\cite{Bergfeld:1994gt,Albrecht:1991bu,pdg} and they have been the object
of theoretical study. Different semileptonic decays of the $\Lambda_c$
have been studied theoretically within the formalism of heavy quark
theory \cite{Korner:1994nh,Richman:1995wm,Gronau:2010if}. On the other
hand, semileptonic decays of $B$ mesons have got comparatively more
attention, both experimentally and theoretically. Quark models are used
for $B$ and $D$ decays in \cite{Isgur:1988gb}, heavy quark effective
theory is also considered in \cite{Manohar:1993qn}, the light-front
formalism is used in a relativistic calculation of form factors for
semileptonic decays in the constituent quark model in \cite{Jaus:1989au}
and lattice QCD calculations have also brought their share to this
problem \cite{Dalgic:2006dt}. More recently the issue has been
retaken. Quark models for semileptonic decays are used in
\cite{Albertus:2011xz,Albertus:2012jt,Albertus:2014xwa}, heavy meson
chiral perturbation theory is found particularly suited in case with
small recoil (when the lepton pair carries much energy) and is used in 
\cite{Kang:2013jaa}, while for large recoil, an approach that combines both hard-scattering and
low-energy interactions has been developed in \cite{Meissner:2013hya}.

    Particular interest is offered by semileptonic decays into a pair of mesons when this pair interacts strongly giving rise to resonances. The interest then is focused on the region of invariant masses where the resonance appears, looking for different channels. The fact that one needs only to study a narrow window of invariant masses allows one to use the practically constant hard form factors of these processes in that range and concentrate on the effects of the meson meson interaction, thus learning about details of hadron interactions, and eventually of the nature of the resonances that are formed in the process.  This is the spirit of the works 
\cite{navarra,semiseki}. In \cite{navarra} the molecular nature of the $D_{s0}^*(2317)$ and $D_0^*(2400)$ resonances is tested using the semileptonic $B_s$ and $B$ decays. In \cite{semiseki} the nature of the light scalar mesons $f_0(500)$, $f_0(980)$, $a_0(980)$ and $\kappa(800)$ is tested  with the semileptonic decays of $D$ mesons. 

 The $\Lambda(1405)$ is an emblematic baryon resonance which has captured the attention of hadron physicists for long, because it does not follow the standard pattern of the three quark baryons. Indeed, in \cite{Dalitz:1960du,Dalitz:1967fp} it was already suggested that 
this resonance should be a molecular state of $\bar K N$ and $\pi \Sigma$. The advent of  chiral unitary theory has allowed to make this idea more precise and consistent with the basic dynamics of QCD encoded in the chiral Lagrangians \cite{Kaiser:1995eg,Kaiser:1996js,Oset:1998it,ollerulf,Lutz:2001yb,
Oset:2001cn,Hyodo:2002pk,cola,GarciaRecio:2002td,GarciaRecio:2005hy,Borasoy:2005ie,
Oller:2006jw,Borasoy:2006sr,hyodonew,Mai:2014xna}. Early in the developments of this theory it was found in \cite{ollerulf} that there are two poles in the same Riemann sheet and, hence, two states, associated to this resonance. A detailed study was done in \cite{cola} by 
looking at the breaking of SU(3), which confirmed the existence of 
these two poles and its dynamical origin. One of the consequences of the existence of these two states is that the shape of the resonance varies from one reaction to another, depending on the weight given to each one the poles by the reaction mechanisms \cite{Thomas:1973uh,Hemingway:1984pz,Niiyama:2008rt,prakhov,Moriya:2012zz,Moriya:2013eb, Zychor:2007gf,fabbietti}. Originally most reactions showed peaks around 1400~MeV, 
from where the nominal mass of the resonance was taken, but the $K^- p \to \pi^0\pi^0 \Sigma^0$
 \cite{prakhov} showed a neat peak around 1420~MeV, 
narrower than the one observed in \cite{Thomas:1973uh,Hemingway:1984pz}. This feature was interpreted within the chiral unitary approach in \cite{magaslam}, showing the mechanisms that gave bigger weight to the resonance with higher mass. Another 
relevant experiment was the one of \cite{Braun:1977wd} which showed a 
peak around 1420~MeV in the $K^- d \to n \pi \Sigma$ reaction, 
also interpreted theoretically in \cite{sekihara} along the same lines (see also the related Refs.~\cite{Miyagawa:2012xz,Jido:2012cy}). Similarly, the thorough data on $\pi \Sigma$ photoproduction at Jefferson Lab \cite{Moriya:2012zz,Moriya:2013eb}, and the subsequent analysis in \cite{Roca:2013av,Roca:2013cca}, has further clarified the situation concerning the two $\Lambda(1405)$ states. 

The need to further clarify the existence and nature of these two $\Lambda(1405)$ states has prompted the suggestion of new reactions, using weak decays processes which, due to particular selection rules, act as filters of isospin I=0 and allow the formation of the  $\Lambda(1405)$ without contamination of the I=1. This is the case of the $\Lambda_b \rightarrow J/\psi ~ \Lambda (1405)$ decay proposed in \cite{Roca:2015tea}, which has partially been measured in \cite{Aaij:2015tga} and the $\Lambda_c \rightarrow \pi ~ \Lambda (1405)$ proposed in \cite{Miyahara:2015cja} and currently under study at Belle \cite{suzuki}. The
neutrino induced production 
of the $\Lambda(1405)$ has also been suggested as a good tool to investigate the properties and 
nature of this resonance \cite{Ren:2015bsa}. 

  In the present work we study theoretically the semileptonic $\Lambda_c$ decay to $\nu l^+$ and $\Lambda(1405)$, which, as we shall see, is a perfect filter of I=0 and, hence, a very good instrument to isolate the $\Lambda(1405)$ and study its properties. The work combines the findings of $\Lambda_b \rightarrow J/\psi ~ \Lambda (1405)$ decay in \cite{Roca:2015tea} with those of the semileptonic $D$ decay studied in \cite{semiseki}, and makes absolute predictions for invariant mass distributions of $\pi \Sigma$, from where the signal of the $\Lambda(1405)$ should be seen, and $\bar K N$, in the reaction $\Lambda_c \to \nu l^+ MB$, with MB either $\pi^+ \Sigma^-$, $\pi^- \Sigma^+$, $\pi^0 \Sigma^0$, $K^- p$ and $\bar K^0 n$.

\section{Formalism}

\begin{figure}[htb]
\begin{center}
\includegraphics[height=3cm]{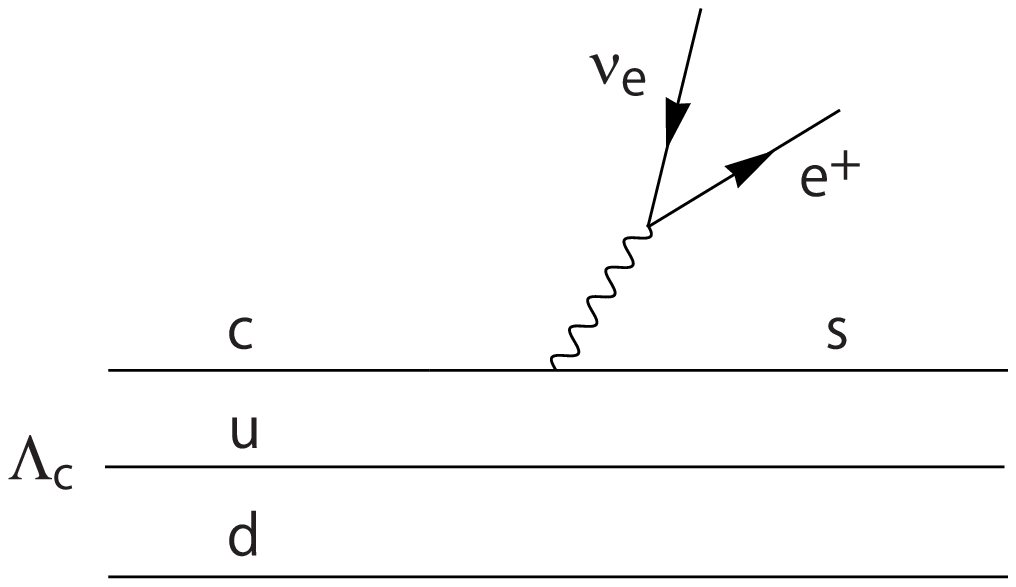}
\caption{Diagrammatic representation of the quark level for $\Lambda_c
 \rightarrow \nu_{e} e^+ (sud)$.}
\label{fig:1}
\end{center}
\end{figure}

The $\Lambda_c \rightarrow \nu e^+ \Lambda(1405)$ process proceeds at the
quark level through a first step shown in Fig.~\ref{fig:1}.

\begin{figure}[htb]
\begin{center}
\includegraphics[height=3.5cm]{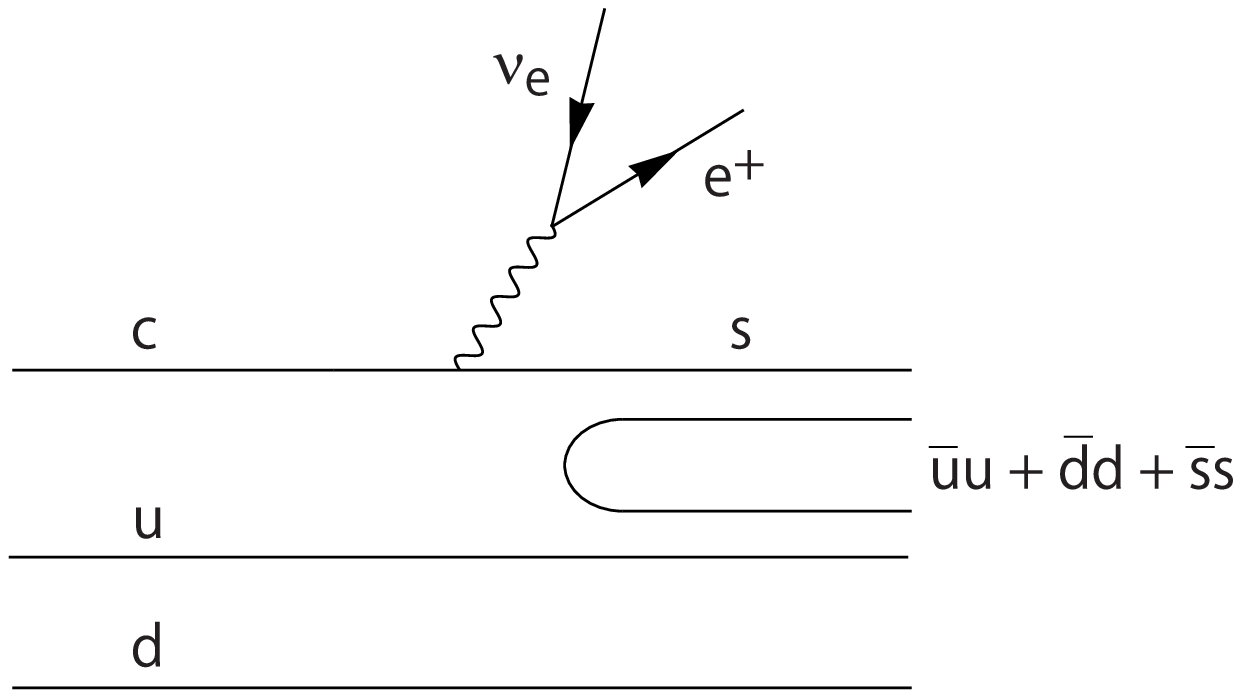}
\caption{Dominant mechanism for the hadronization into meson baryon
 of the $sud$ state after the weak process.}
\label{fig:2}
\end{center}
\end{figure}

The process involves the $cs$ weak transition, which is Cabibbo favored,
and 
this is the same one as in the $D$ decays studied in \cite{semiseki}.
There is, however, a novelty in the present process.
Indeed, if we want to see the $\Lambda(1405)$ this must be done in the
mass distribution of $\pi \Sigma$, hence the $sud$ quark of
Fig.~\ref{fig:1} must hadronize into a meson baryon component.
This is done easily for mesons since one introduces an extra $\bar{q} q$ meson
with vacuum quantum number, $\bar{u}u + \bar{d}d + \bar{s}s$, and then
the two quarks after the weak process participate
in the formation of the two mesons.
With three quarks after the weak vertex, as in Fig.~\ref{fig:1}, the new
$\bar{q} q$ pair can be placed in between different pairs of the $sud$
quarks. However, there is some reason to do that involving the $s$
quark, such that the $s$ quark goes into the emerging meson. 
The reasons are the following:

\begin{itemize}
 \item[1)]  The $\Lambda_c$ has $I=0$, and this forces the $ud$ initial
	   state to be in a $I=0$ state,
	   $\displaystyle \frac{1}{\sqrt{2}}(ud-du)$. The dominant mechanism that
	   requires just a one body operator involving the $c$ and $s$
	   quarks will leave the $ud$ original
	   quarks as spectators and will still have $I=0$ in the final
	   state. They are also in $L=0$ in a quark picture of the wave
	   function. Since the $\Lambda(1405)$ has negative parity, it
	   is the $s$ quark that will convey this parity, hence being in
	   $L=1$ in a quark picture. Since in the $\bar{K}N$, or meson
	   baryon states in general, all quark are in the ground
	   state, the $s$ quark must be deexcited and hence it has to
	   participate in the hadronization.

\item[2)] Even then we have the option to have the $s$ quark belonging
	to the meson or to the baryon. If the $s$ quark goes into the
	meson, the original $u, d$ quarks are spectators and  go into the final
	baryon, the other needed quark coming from the additional
	$\bar{q}q$ of the hadronization. If the $s$ quark goes into the
	baryon, the original $u$ or $d$ quark must go into the
	meson. In these processes the baryon is the heaviest particle
	and the phase space favors the lighter mesons and leptons to
	carry the momenta and, thus, the energy. If we have $\nu_{e}
	l^{+} \pi \Sigma$, the pair $\nu_{e} l^+$ and the pion would
	basically carry all the energy and then the $\pi$ (and the
	$\nu e^+$) has about 550 MeV/c. Also the $s$ quark will carry
	the 
	same momentum as the $\nu_{e} e^{+}$ pair and go into the
	$\Sigma$, essentially at rest. One has then form factors from
	the quark nuclear wave functions, involving twice a fair amount
	of momentum transfer and the mechanism is suppressed versus the
	one where the original $u, d$ quarks are spectators.  
After the former discussion the dominant mechanism for the hadronization
is depicted in Fig.~\ref{fig:2}. This is what was done in the $\Lambda_c
\rightarrow J/\psi M B$~\cite{Roca:2015tea}
and the $\Lambda_c \rightarrow \pi M B$~\cite{Miyahara:2015cja} reactions.
\end{itemize}

\subsection{Hadronization}
The procedure followed here is inspired in the approach of 
\cite{Liang:2014tia} where the basic mechanisms at the quark level are
investigated, then pairs of hadrons are produced after implementing
hadronization, and finally these hadrons are allowed to undergo final
state interaction.
With the $u, d$ quarks as spectators in $I=0$ the final $s u d$ state
also has $I=0$ and we, thus, have a filter of $I=0$.
Then, upon hadronization the final meson baryon is constructed as
follows~\cite{Roca:2015tea},
\begin{eqnarray}
|H \rangle &=& \frac{1}{\sqrt{2}} |s (\bar{u}u + \bar{d}d + \bar{s}s) (ud
 - du)) \rangle  \nonumber\\
&=& \frac{1}{\sqrt{2}} \sum_{i=1}^{3} P_{3i} q_{i}(ud - du) \nonumber
\end{eqnarray}
where
$\displaystyle q \equiv \left\{  
    \begin{array}{c}
      u \\
      d \\
      s 
    \end{array}
\right\} $
and
$\displaystyle P \equiv q \bar{q}^{\tau} = \left( 
\begin{array}{ccc}
 u\bar{u} & u\bar{d} & u\bar{s} \\
 d\bar{u} & d\bar{d} & d\bar{s} \\
 s\bar{u} & s\bar{s} & s\bar{s} 
\end{array}
\right)$.\\
It is convenient to write the $q\bar{q}$ matrix, $P$,
in terms of mesons and we have
\begin{eqnarray}
P \equiv \left( 
\begin{array}{ccc}
\frac{\pi^0}{\sqrt{2}} + \frac{\eta}{\sqrt{3}} +\frac{\eta^{\prime}}{\sqrt{6}} & \pi^+ & K^+ \\
\pi^- & -\frac{\pi^0}{\sqrt{2}}+\frac{\eta}{\sqrt{3}}+\frac{\eta^{\prime}}{\sqrt{6}}  & K^0 \\
K^- & \bar{K}^0 & -\frac{\eta}{\sqrt{3}}+\frac{2\eta^{\prime}}{\sqrt{6}} 
\end{array}
\right)
\nonumber
\end{eqnarray}
where the standard $\eta, \eta^{\prime}$ mixing has been
assumed~\cite{Roca:2015tea,Bramon:1992kr}.
Neglecting the $\eta^{\prime}$ terms because of its large mass, as
done in \cite{Roca:2015tea},
we have
\begin{eqnarray}
|H \rangle &=& \frac{1}{\sqrt{2}}(K^{-} u(ud-du) + \bar{K}^{0} d(ud - du)  \nonumber\\
& & - \frac{1}{\sqrt{3}}\eta s(ud -du) ). 
\nonumber
\end{eqnarray}
We can see that we have now the mixed antisymmetric representation of
the octet of baryons.
By taking this wave function for the $p, n$ and $\Lambda$ states 
(see \cite{close}) we find \cite{Roca:2015tea}
\begin{eqnarray}
|H \rangle = | K^{-} p \rangle + | \bar{K}^0 n \rangle -
 \frac{\sqrt{2}}{3} | \eta \Lambda \rangle. 
\label{eq:H}
\end{eqnarray}
One must recall that the $\Lambda(1405)$ is obtained in coupled channel
from the interaction of the $\bar{K}N, \pi\Sigma, \eta\Lambda, K\Xi$
states, however, as a first step only the $\bar{K}N$ and $\eta\Lambda$ 
states are formed, but not the $\pi\Sigma$ and $K\Xi$.
Since the $\Lambda(1405)$ is seen in the $\pi\Sigma$ invariant mass spectrum,
the only way to see $\pi\Sigma$ is through rescattering of the $\bar{K}N$
and $\eta\Lambda$ states and this involves directly the transition
$\bar{K}N \rightarrow \pi\Sigma$ and $\eta\Lambda \rightarrow \pi\Sigma$ matrix elements that
contain the $\Lambda(1405)$ pole.
In other words, we do not have $\pi\Sigma$ at the tree level and,
thus, one avoids background, hence, stressing more the $\Lambda(1405)$ signal. 
We have these two benefits in the reaction: there is an $I=0$ filter, and
the $\Lambda(1405)$ is produced with no background.

\subsection{Final state interaction}
\begin{figure}[htb]
\begin{center}
\includegraphics[width=8cm]{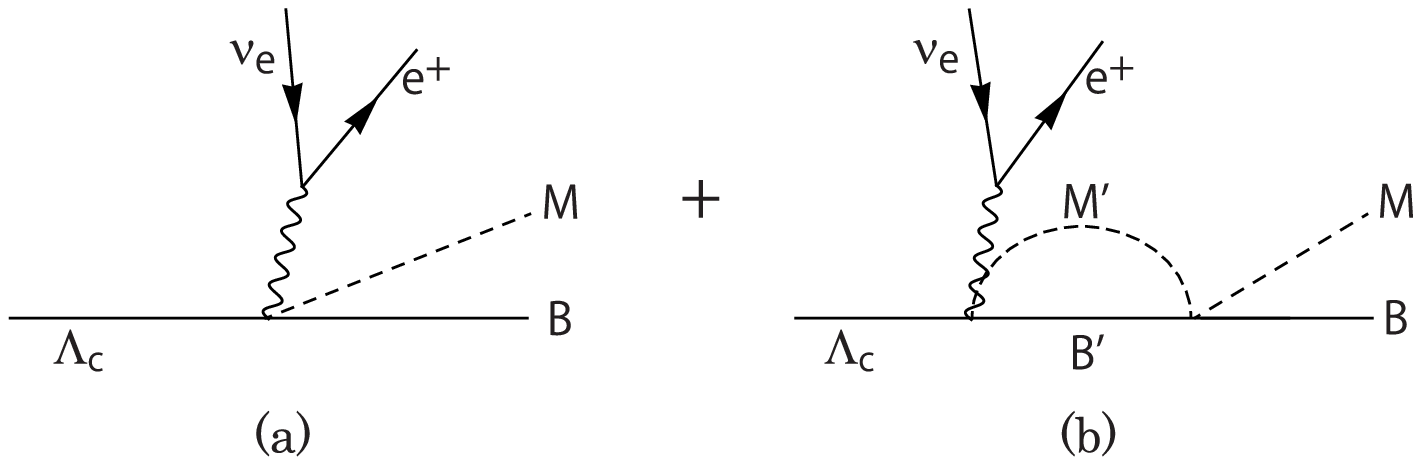}
\caption{Diagrams involved in the final state interaction of the primary
 $MB$ mesons, (a) tree level, (b) rescattering.}
\label{fig:3}
\end{center}
\end{figure}

The final state interaction is depicted in Fig.~\ref{fig:3}.

The matrix element for the $\Lambda_c \rightarrow (\nu_{e} e^{+}) M_j B_j$
is then given by
\begin{equation}
 t_{{\rm had}, {j}}(M_{\rm inv}) = C \left( 
h_{j} + \sum_{i} h_{i}
  G_{i}(M_{\rm inv}) t_{ij}(M_{\rm inv}) \right)
\label{eq:t}
\end{equation}
where from Eq.~(\ref{eq:H}) we have 
\begin{eqnarray}
h_{\pi^0 \Sigma^0} &=& h_{\pi^+ \Sigma^-} = h_{\pi^- \Sigma^+} =  0, 
\hspace{5mm}
h_{\eta\Lambda} = -\frac{\sqrt{2}}{3}, \nonumber\\  
h_{K^- p} &=& h_{\bar{K}^0 n}=1,
\hspace{5mm}
h_{K^+ \Xi^-} = h_{K^0 \Xi^0} = 0,
\nonumber\\
h_{\pi^0 \Lambda} &=& h_{\eta \Sigma^0} = 0.
\nonumber
\end{eqnarray}

In Eq.~(\ref{eq:t}) $G_{i}$ is the loop function of meson baryon and $t_{ij}$
the scattering matrix in the basis of states $K^- p$, $\bar{K}^0 n$,
$\pi^0\Lambda$, $\pi^0\Sigma^0$, $\eta\Lambda$, $\eta\Sigma^0$, $\pi^+\Sigma^-$,
$\pi^-\Sigma^+$, $K^+\Xi^-$, $K^0\Xi^0$ with $t$ given by the
Bethe-Salpeter equation 
\begin{equation}
 t=[1 - VG]^{-1} V
\label{eq:BS}
\end{equation}
and $V$ the transition potential taken from \cite{Oset:1998it}.
The $G$ function is regularized with a cut off in three momenta and we take
$q_{\rm max} = 630$~MeV as in \cite{Oset:1998it}.
The factor $C$, which we take constant in the limited range of $M_{\rm inv}$
that we will study, 
encodes matrix element of the hadronization.

\subsection{The weak vertex}
We must take into account the weak vertex and altogether the matrix
element $T_{\Lambda_{c}}$ for the semileptonic decay is given by \cite{semiseki}
\begin{equation}
 T_{\Lambda_c} = \frac{G_{\rm F}}{\sqrt{2}}L^{\alpha}Q_{\alpha} t_{\rm had}
\end{equation}
where
\begin{equation}
 L_\alpha = \bar{u}_\nu \gamma^\alpha (1 - \gamma_5) v_l, 
\hspace{3mm}
 Q_\alpha = \bar{u}_q \gamma_\alpha (1 - \gamma_5) u_c.
\end{equation}
By following the steps of \cite{navarra,semiseki} we find for the sum
and average over the polarization of the fermions
\begin{eqnarray}
 \frac{1}{2} \sum_{\rm pol} |T_{\Lambda_c}|^2 = \frac{ 4 |G_{\rm F} t_{\rm had}
  V_{cs}|^2}
{m_e m_\nu m_{\Lambda_c} m_R}
(p_{\Lambda_c} \cdot p_\nu)(p_R \cdot p_e)
\end{eqnarray}
where $R$ stands for the final $MB$ system formed and, thus, $M_R = M_{\rm
inv}$. Further steps are done in \cite{navarra} to perform the angular
integration of the leptons in the $\nu e$ rest frame and 
finally one obtains a formula for $\displaystyle \frac{d\Gamma}{dM_{\rm inv}}$ 
given by
\begin{eqnarray}
 \frac{d \Gamma_{i}}{dM_{\rm inv}} &=& \frac{|G_{\rm F} V_{cs} t_{{\rm had},i}|^2}
{32 \pi^5 m_{\Lambda_c}^3 M_{\rm inv}^{(i)} } \nonumber\\
&\times&
\int dM_{\rm inv}^{(\nu e)} P^{\rm cm} \tilde{p}_{\nu} \tilde{p}_{i}
( M_{\rm inv}^{(\nu e)} )^2 \left( \tilde{E}_{\Lambda_c} \tilde{E}_{i} - \frac{\tilde{p}_{\Lambda_c}^2}{3} \right) 
\nonumber\\
\label{eq:dGam}
\end{eqnarray}
where $P^{\rm cm}$ is the momentum of the $\nu e$ system in the $\Lambda_c$
rest frame, $\tilde{p}_{\nu}$ is the momentum of the neutrino in the
$\nu e$ rest frame, $\tilde{p}_{i}$ is the relative momentum of the
final meson in the $(MB)_i$ rest frame, $\tilde{E}_{\Lambda_c}$,
$\tilde{E}_i$ the $\Lambda_c$ and $(MB)_i$ energy in the $\nu e$ rest
frame and $\tilde{p}_{\Lambda_c}$ the momentum of the $\Lambda_c$ in the
$\nu_e$ rest frame;
\begin{eqnarray}
P^{\rm cm} &=& \frac{\lambda^{1/2}(m_{\Lambda_c}^2, M_{\rm inv}^{(\nu e) 2},
  M_{\rm inv}^{(i) 2} )}{2m_{\Lambda_c} },  \\
\tilde{p}_{\nu} &=& \frac{\lambda^{1/2}(M_{\rm inv}^{(\nu e) 2}, m_{\nu}^2,
  m_{e}^2 )}{2M_{\rm inv}^{(\nu e)} },  \\
\tilde{p}_{i} &=& \frac{\lambda^{1/2}(M_{\rm inv}^{(i) 2}, m_{i}^2,
  M_{i}^2 )}{2M_{\rm inv}^{(i)} }, 
\end{eqnarray}
with $m_i$, $M_i$ the meson, baryon masses of the $(MB)_i$ final state,
\begin{eqnarray}
\tilde{E}_{\Lambda_c} &=& \frac{m_{\Lambda_c}^2 + M_{\rm inv}^{(\nu e) 2} -
 M_{\rm inv}^{(i)2} }{2 M_{\rm inv}^{(\nu e)} }, \\
\tilde{E}_{i} &=& \frac{m_{\Lambda_c}^2 - M_{\rm inv}^{(\nu e) 2} -
 M_{\rm inv}^{(i)2} }{2 M_{\rm inv}^{(\nu e)} },\\
\tilde{p}_{\Lambda_c}^{2} &=& \tilde{E}_{\Lambda_c}^2 - m_{\Lambda_c}^2.
\end{eqnarray}
We take $G_{\rm F} = 1.166 \times 10^{-5}$ GeV$^{-2}$, 
$(V_{cs}) = \cos\theta_c = 0.986$
and for the constant $C$ we take the same value that was obtained in the
semileptonic decay of $D$ to two mesons~\cite{semiseki}, which involves
the $cs$ transitions as here, $C=4.597$ was adjusted to experimental
data of the $D^+ \rightarrow (\pi^+ K^-)_{s- {\rm wave}} e^+ \nu_e$ reaction.
This is not necessarily the case, since in \cite{semiseki} one studied
decay of mesons and here we have decay of baryons.
The matrix elements are not necessarily the same, but as an order of
magnitude this can serve.
The real predictions of the work are the shape of the $\Lambda(1405)$, 
which is due to the state around 1420 MeV, and the relative strength of
$\bar{K}N$ mass distribution above the $\bar{K}N$ threshold and the
$\pi\Sigma$ distribution at the resonance peak.
As to the absolute ratio, we can have a feeling for the uncertainties by
taking also the value of $C=7.22$ obtained from the semileptonic $B$
decays in \cite{navarra}.

\section{Results}
\begin{figure}[tb]
\begin{center}
\includegraphics[width=7.5cm]{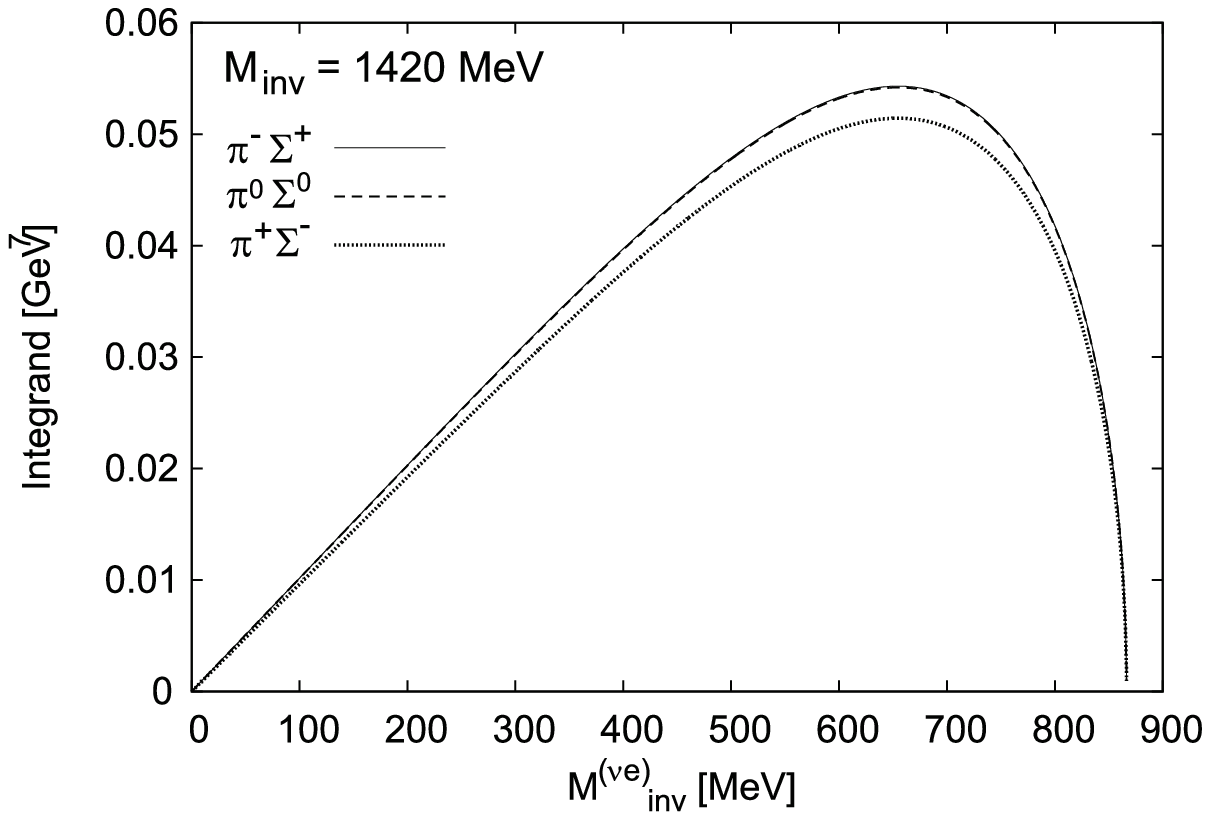}
\includegraphics[width=7.5cm]{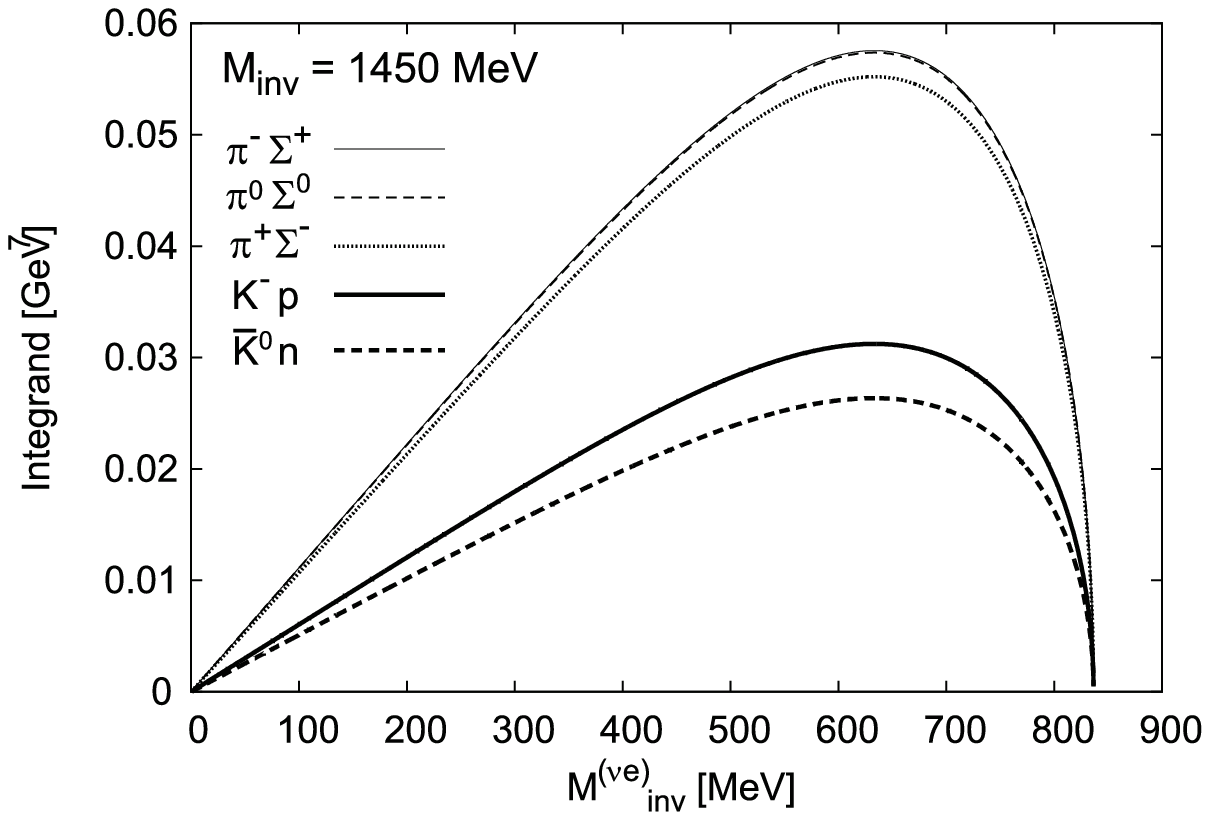}
\includegraphics[width=7.5cm]{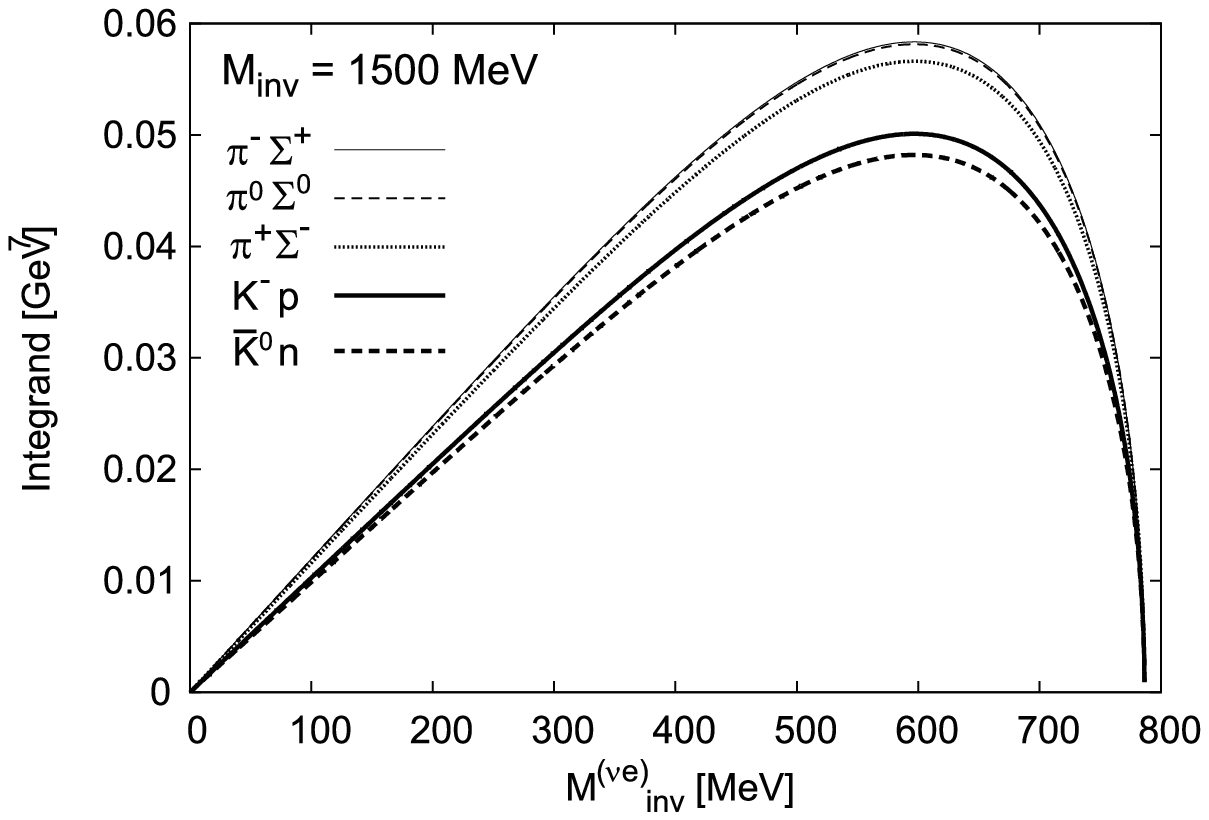}
\caption{The integrand of the integral that appear
Eq.~(\ref{eq:dGam}) as a function $M_{\rm inv}^{(\nu e)}$ for different values
 of the invariant mass of the final $MB$ pair.
The value $C=4.597$ is used.}
\label{fig:4}
\end{center}
\end{figure}

In Fig.~\ref{fig:4} we plot the integrand of the integral that appears
Eq.~(\ref{eq:dGam}) for different values of $M_{\rm inv}$.
The calculations are done for $\nu_e e^+$.
The results for $\nu_{\mu} \mu^+$ are very similar. 
As we can see, the strength of this distribution peaks at large values of
$M_{\rm inv}^{(\nu e)}$, close to its maximum value, something that we
already anticipated and used in discussions in the former section.
This justifies taking elements of the work of \cite{Kang:2013jaa}, 
where the $s$-wave form factors prior to final state interaction of the
hadrons are found very smooth, which justifies out choice of a
constant $C$ in our limited range of invariant masses.

\begin{figure}[tb]
\begin{center}
\includegraphics[width=8.5cm]{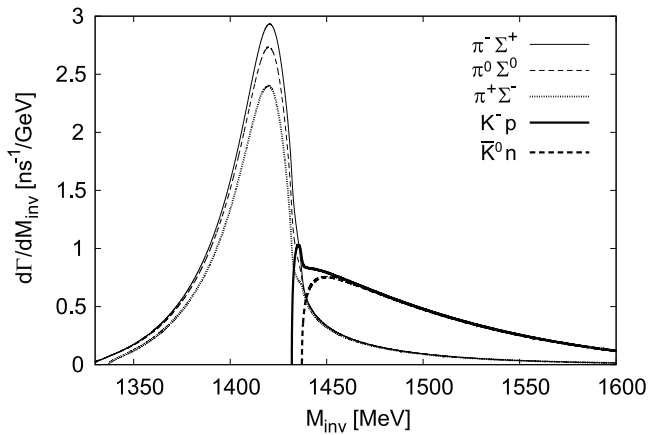}
\caption{The invariant mass distributions of Eq.~(\ref{eq:dGam}) for
 different channels.
The value $C=4.597$ is used.}
\label{fig:5}
\end{center}
\end{figure}

In Fig.~\ref{fig:5} we show the final result for the invariant mass
distribution of Eq.~(\ref{eq:dGam}) for $\pi^+ \Sigma^-$, 
$\pi^- \Sigma^+$, $\pi^0 \Sigma^0$, $K^- p$ and $\bar{K}^0 n$ in the
final state. We observe neat peaks for $\pi \Sigma$ around 1420~MeV. 
This means that one is mostly exciting the $\Lambda(1405)$ state at 1420
MeV. The other state is around 1385~MeV, but does not play much of a role
in this reaction. This has a dynamical reason. 
As is well known from the chiral unitary approach, the high mass
$\Lambda(1405)$ state couples mostly to $\bar{K}N$, while the one at
1385 MeV couples more strongly to $\pi \Sigma$. Since in the primary
production of $MB$ we produce $\bar{K} N$, but not $\pi \Sigma$, it
becomes clear that the resonance excited is the one around 1420 MeV.
This is a neat prediction of the chiral unitary approach for these
resonances that the experiment could prove or disprove.

We observe some differences in the strength for $\pi^- \Sigma^+$, 
$\pi^+ \Sigma^-$ and $\pi^0 \Sigma^0$.
This is because, even if we have filtered $I=0$ in this reaction, the
scattering matrices induce a bit of isospin breaking became the different
masses of mesons and baryons in the same isospin multiplet.
This can also be observed in the Bonn model in Fig.~3 of \cite{Roca:2015tea}.
The result and the ordering of the strength of the channels is
remarkable similar to what is obtained in
Fig.~5 of \cite{Miyahara:2015cja}.
 
The strength for the $K^- p$ and $\bar{K}^0 n$ production is also
similar to what is obtained in ~\cite{Roca:2015tea} and \cite{Miyahara:2015cja}.
The strength of the $\bar{K}N$ distribution at its peak is about one
fourth of the strength of the $\pi \Sigma$ distribution at its peak, 
a feature also shared by these different works, and this is also a
prediction of the chiral unitary approach.
The fall down with $M_{\rm inv}$ of this distribution depends 
somewhat on the reaction and the particular model used. 
Here we use the lowest order Weinberg Tomozawa term for the kernel $V$
in Eq.~(\ref{eq:BS}), while in \cite{Roca:2015tea}, in the Bonn model,
higher order terms were considered in the kernel~\cite{Mai:2014xna}. 

The values obtained for $d\Gamma/dM_{\rm inv}$ are of the same order of
magnitude as those found in \cite{semiseki} for the $D$ semileptonic
decays with two mesons in the final state.

If we integrate the strength below the $\Lambda(1405)$ peak, we find
$\Gamma \simeq 0.108$ ns$^{-1}$ and the mean life of the $\Lambda_c$
is $5 \times 10^3$ ns$^{-1}$.
Thus we are taking about a branching ratio of about $2 \times 10^{-5}$,
which is within measuring range, since boundaries of $10^{-6}$ for
certain branching ratios have been stablished~\cite{pdg}.
If we use instead the vale $C=7.22$ from \cite{navarra}, then the
branching ratio becomes $5 \times 10^{-5}$.
Accepting uncertainties in the value of $C$, the message is that the ratios obtained are within
present measurable capacities.

\section{Conclusions}

We have studied the semileptonic decay of the $\Lambda_c \to \nu_l l^+ MB$ with MB, a pair of meson baryons, $\pi \Sigma$ or $\bar K N$. The idea is to look for the $\pi \Sigma$ mass distribution where the $\Lambda(1405)$ state of higher energy (around 1420 MeV) should show up. The reaction was shown to have some welcome features: it filters I=0 and thus one avoids having to separate contributions from I=1 which complicate the analysis of data in other reactions. Next, since the primary production of meson baryon in this process does not produce $\pi \Sigma$, this channel will appear through rescattering of the $\bar K N$ and $\eta \Lambda$ which are the states primarily produced. This avoids background originated from tree level terms and the resonance shows up more clearly. Finally, the chiral unitary approach predicts two states around the $\Lambda(1405)$, the one with lower mass coupling mostly to $\pi \Sigma$, and the one of higher mass coupling mostly to $\bar K N$.  Since the reaction is such that the resonance is produced though rescattering of a primary $\bar K N$, this guarantees that one will see mostly this latter state and we predict a clean peak around 1420 MeV, that the experiment could support or disprove. In addition we also make predictions for production of $\bar K N$ pairs, which are tied to the dynamics of the chiral unitary approach and could be investigated at the same time. 

     The fact that we only need the mass distribution in a narrow region of the invariant mass, allows us to factorize in a constant all elements of the weak transition and the hadronization which are not explicitly considered, and the dependence of the final distributions on the invariant mass is then tied to the final state interaction of the meson baryon pairs, which in particular generate the two $\Lambda(1405)$ states. 

While most of the theoretical hadron community would agree on the existence of two states of the $\Lambda(1405)$ and the molecular nature of these states, there are still some differences as to the position of the lower mass state and details on the mass distributions. Also, these ideas are not so broadly accepted in the experimental hadron community. It is most convenient to carry out these experiments where neat predictions are made which are tied to the concrete chiral dynamics in coupled channels that generates these states. Experimental confirmation of these predictions would be a step forward to consolidate these ideas and make progress in the understanding of the nature of baryons.

\section*{Acknowledgments}

One of us, N. I., wishes to acknowledge
the support by Open Partnership Joint Projects of JSPS Bilateral Joint Research Projects.
This work is partly supported by the Spanish Ministerio de Economia y
Competitividad and European FEDER funds under the contract number
FIS2011-28853-C02-01 and FIS2011-28853-C02-02, and the Generalitat
Valenciana in the program Prometeo II-2014/068. We acknowledge the
support of the European Community-Research Infrastructure
Integrating Activity Study of Strongly Interacting Matter (acronym
HadronPhysics3, Grant Agreement n. 283286) under the Seventh
Framework Programme of EU. .

\bibliographystyle{plain}

\end{document}